\documentclass[12pt,a4paper]{article}
\usepackage{graphicx}
\usepackage[caption=false]{subfig}
\usepackage{epsfig}
\usepackage{amssymb}
\usepackage{amsmath}
\usepackage{lipsum}
\usepackage{subfig}
\usepackage[left=2cm,right=2cm,top=2cm,bottom=2cm]{geometry}
\usepackage{multirow}
\usepackage{amsmath}
\usepackage[toc]{appendix}
\begin{document}
\title{Isgur-Wise function and a new approach to potential model}
\author{Tapashi Das$^{1,\ast}$, D. K. Choudhury$^{1,2}$ and K. K. Pathak$^{3}$\\  $^1$Department of Physics,\ Gauhati University,
Guwahati-781014, India.\\
$^2$ Physics Academy of North-East, Guwahati-781014, India. \\
$^3$ Department of Physics, Arya Vidyapeeth College, Guwahati-781016, India\\
$^\ast$\small{email:t4tapashi@gmail.com}}
\date{}
\maketitle
\begin{abstract} 
Considering the Cornell potential $V(r)=-\frac{4\alpha_s}{3r}+br+c$, we have revisited the Dalgarno's method of perturbation by incorporating two scales $r^{short}$ and  $r^{long}$ as integration limit so that the perturbative procedure can be improved in a potential model. With the improved version of the wave function the ground state masses of the heavy-light mesons $ D, D_s, B, B_s$ and $B_c $ are computed. The slopes and curvatures of the form factors of semi-leptonic decays of heavy-light mesons in both HQET limit and finite mass limit are calculated and compared with the available data.
\end{abstract}



\section{Introduction}
The heavy hadron spectroscopy played a major role in the foundation of QCD. In last few years, it has sparked a renewed interest in the subject due to numerous data available from the B factories\cite{bfactory}, CLEO \cite{cleo}, LHCb \cite{lhcb} and the Tevatron \cite{1}. In more recent times the discovery of X-Y states \cite{a} as possible charmonium and bottonium hybrids have extended such study of the exotic heavy hadron spectroscopy. The most recent discoveries of the charmonium pentaquarks \cite{b} have further increase its importance. The simplest system of this area are the heavy-light and heavy-heavy hadrons.\\

In the present paper, we will report a study of such heavy flavored mesons in a QCD potential model \cite{potential model} persued in recent years. In the last few years, the experimental study of heavy-light and heavy-heavy mesons have renewed the theoretical interest towards HQET (Heavy Quark Effective Theory) and Isgur-Wise function \cite{prd90,epjc14,kkbj,5,h}.\\

The dynamics of the heavy quark meson is governed by the inter-quark potential. The properties of the heavy mesons are in rough approximation is described by the Cornell potential, $V(r)=-\frac{4\alpha_s}{3r}+br+c$ \cite{cornell}, which is a Coulomb-plus-linear non-relativistic confinement potential. The first Coulomb term of the potential is consistent with one-gluon-exchange contribution for short distance. The second term generates the confinement in long distance. Both the potentials play decisive role in the quark dynamics and their seperation is not possible. Besides there is no appropriate small parameter so that one of the potential within a perturbative theory can be made perturbative. The third term `$c$' \cite{yang} which is a phenomenological constant needed to reproduce correct masses of heavy-light meson bound state.\\

In general, it is expected that a constant term `$c$' in the potential should not affect the wave function of the system while applying the perturbation theory. But in our previous work \cite{ijp} it is seen that whether the term `$c$' is in parent or perturbed part of the Hamiltonian, it always appears in the total wave function which is inconsistent with the quantum mechanical idea that a constant term `$c$' in the potential can at best shift the energy scale, but should not perturb the wave function i.e. a Hamiltonian $H$ with such a constant and another $H^\prime$ without it should give rise to the same wave functions.\\

Due to this inconsistency or for the validation of the quantum mechanical idea while using perturbation theory like Dalgarno's method \cite{2,3} in the present work we have considered the scaling factor $c=0$.\\

Also in the present work both the short range and long range effect is tried to incorporate in the total wave function. Because in our earlier works \cite{5,pd,13}, the properties of the mesons are studied considering the Coulombic part of the Cornell potential dominant over the linear part. On the other hand in ref. \cite{kkbj,bjdk,6}, the Schrodinger equation is solved by considering the linear part to be dominant over the Coulombic part.\\

However it is well known that at short distance Coulomb potential plays a more dominant role than the linear confinement because while the former is inversely proportional to `$r$', the later is linear. Similarly, for large distance the confinement takes over the Coulomb effect. Therefore if the inter-quark seperation `$r$' can be roughly divided into short distance ($r^{short}$) and long distance ($r^{long}$) effectively one of the potential will dominate over the other. In such situation confinement parameter ($b$) and the strong coupling parameter ($\alpha_s$) can be considered as effective and appropriate small pertubative parameters.\\

The present paper is organised as follows: in section 2, we outline the formalism, while in section 3 summarize the results for masses of various mesons and slope and curvature for Isgur-Wise function. Section 4 contains conclusion and comments. 

\section{Formalism}
\subsection{Dalgarno's method of perturbation:}
The non-relativistic two body Schrodinger equation \cite{3} is
\begin{equation}
H|\psi\rangle=(H_0+H^\prime)|\psi\rangle=E|\psi\rangle,
\end{equation}

so that the first order perturbed eigenfunction $\psi^{(1)}$ and eigenenergy $W^{(1)}$ can be obtained using the relation
\begin{equation}	
H_0 \psi^{(1)} + H^\prime \psi^{(0)}=W^{(0)}\psi^{(1)} + W^{(1)} \psi^{(0)},
\end{equation}

where
\begin{equation}
W^{(0)}=  <\psi^{(0)}\vert H_0 \vert \psi^{(0)}>,
\end{equation}
\begin{equation}
W^{(1)}= <\psi^{(0)}\vert H^\prime \vert \psi^{(0)}>.
\end{equation}

We calculate the total wave functions using Dalgarno's method of perturbation for the potential 
\begin{equation}
V(r)=-\frac{4\alpha_s}{3r}+br,
\end{equation}

where -$\frac{4}{3}$ is due to the color factor, $\alpha_s$ is the strong coupling constant, $r$ is the inter-quark distance, $b$ is the confinement parameter (phenomenologically, $b=0.183 GeV^2$ \cite{4}).\\

For potential of type (5), one of the choice for parent and perturbed Hamiltonian is

\begin{equation}
H_0=-\frac{4\alpha_s}{3r} \nonumber
\end{equation}  

and 
\begin{equation}
H^\prime=br \nonumber.
\end{equation}

The total wave function (Appendix-A) for this case is 

\begin{equation}
\psi^{total}_I(r)=\frac{N}{\sqrt{\pi a_0^3}}\left[ 1-\frac{1}{2}\mu b a_0r^2\right] \left( \frac{r}{a_0}\right) ^{-\epsilon}e^{-\frac{r}{a_0}},
\end{equation}

where normalisation constant 

\begin{equation}
N=\frac{1}{\left[ \int_0^{r^{short}} \frac{4 r^2}{a_0^3}\left[ 1-\frac{1}{2}\mu b a_0r^2\right]^2\left( \frac{r}{a_0}\right) ^{-2\epsilon} e^{-\frac{2r}{a_0}}dr\right] ^{\frac{1}{2}}},
\end{equation}

where the cut off parameter $r^{short}$ is used as integration limit for Coulomb as parent and linear as perturbation. Because here Coulomb part is considered to be dominant over the linear part for short distance and

\begin{equation}
a_0=\left( \frac{4}{3}\mu\alpha_s\right)^{-1},
\end{equation}

\begin{equation}
\mu=\frac{m_qm_Q}{m_q+m_Q},
\end{equation}

$m_q$ and $m_Q$ are the masses of the light and heavy quark/antiquark respectively and $\mu$ is the reduced mass of the mesons and

\begin{equation}
\epsilon=1-\sqrt{1-\left( \frac{4}{3}\alpha_s\right) ^2}
\end{equation}

is the correction for relativistic effect \cite{re1,re2} due to Dirac modification factor.\\

Similarly, the wave function upto $O(r^4)$ (Appendix-B) for another choice of parent and perturbed Hamiltonian of (5), \\

where

\begin{equation}
H_0=br
\end{equation}

and

\begin{equation}
H^\prime=-\frac{4\alpha_s}{3r}
\end{equation}

is

\begin{equation}
\psi^{total}_{II}(r)=\frac{N^\prime}{r} \left[1+A_0r^0+A_1(r)r+A_2(r)r^2+A_3(r)r^3+A_4(r)r^4\right]A_i[\rho_1 r+\rho_0] \left( \frac{r}{a_0}\right) ^{-\epsilon},
\end{equation}

where $A_i[r]$ is the Airy function \cite{7} and $N^\prime$ is the normalization constant,
\begin{equation}
N^\prime= \frac{1}{\left[ \int_{r^{long}}^{r_0} 4 \pi \left[1+A_0r^0+A_1(r)r+A_2(r)r^2+A_3(r)r^3+A_4(r)r^4\right]^2 \left( A_i[\rho_1 r+\rho_0]\right) ^2 \left( \frac{r}{a_0}\right) ^{-2\epsilon}dr\right]^{\frac{1}{2}} }.
\end{equation}

The cut off parameter $r^{long}$ is used as integration limit because we have considered linear as parent and Coulomb as perturbation, where the linear part is considered to be dominant over the Coulomb part for long distance. The upper cut off $r_0$ is used to make the analysis normalizable and convergent, because we have used Airy function as meson wave function. Later we fixed $r_0$ to 1 $Fermi$ \cite{bali} for our calculations.\\

The co-efficients of the series solution as occured in Dalgarno's method of perturbation, are the function of $\alpha_s$, $\mu$, and $b$:

\begin{equation}
A_0=0,
\end{equation}
\begin{equation}
A_1=\frac{-2\mu \frac{4\alpha_s}{3}}{2\rho_1 k_1+\rho_1^2 k_2},
\end{equation}

\begin{equation}
A_2=\frac{-2\mu W^1}{2+4 \rho_1 k_1+ \rho_1^2 k_2},
\end{equation}

\begin{equation}
A_3=\frac{-2\mu W^0 A_1}{6+6 \rho_1 k_1+ \rho_1^2 k_2},
\end{equation}

\begin{equation}
A_4=\frac{-2\mu W^0 A_2+2\mu b A_1}{12+8 \rho_1 k_1+ \rho_1^2 k_2}.
\end{equation}

The parameters:
\begin{equation}
\rho_1=(2\mu b)^{\frac{1}{3}},
\end{equation}

\begin{equation}
\rho_0=-\left[ \frac{3\pi (4n-1)}{8}\right] ^{\frac{2}{3}},
\end{equation}

(in our case n=1 for ground state)
\begin{equation}
k=\frac{0.355-(0.258) \rho_0}{(0.258) \rho_1},
\end{equation} 

\begin{equation}
k_1=1+\frac{k}{r},
\end{equation}

\begin{equation}
k_2=\frac{k^2}{r^2},
\end{equation}

\begin{equation}
W^1=\int \psi^{(0)\star} H^{\prime} \psi^{(0)} d\tau,
\end{equation}

\begin{equation}
W^0=\int \psi^{(0)\star} H_0 \psi^{(0)} d\tau.
\end{equation}

\subsection{Ground state masses of mesons}

Masses of heavy flavored mesons in a specific potential model in the ground state can be obtained as:

\begin{equation}
M_P=m_{q/Q}+m_{\bar{q}/\bar{Q}}+\langle H \rangle
\end{equation}

where $m_{q/Q}$ is mass of light (or heavy) quark and $m_{\bar{q}/\bar{Q}}$ is mass of light (or heavy) anti-quark constituting the meson bound state.\\

The above expression shows that to calculate the masses of mesons one needs to find $\langle H \rangle$, so that
\begin{equation}
\langle H \rangle=\langle \frac{p^2}{2\mu}\rangle+ \langle V(r) \rangle \nonumber
\end{equation}

\begin{equation}
=4\pi \int_0^\infty r^2\psi^\ast(r) H \psi(r) dr  \nonumber
\end{equation}

\begin{equation}
=4\pi\int_0^\infty r^2\psi^\ast(r) \left( \frac{p^2}{2\mu}+ V(r)\right)  \psi(r) dr.
\end{equation}

To take into account both the Coulomb and linear part of the potential we improve the above equation with the cut off scales $r^{short}$ and $r^{long}$ as

\begin{equation}
\langle H \rangle = 4\pi\left[ \int_0^{r^{short}} r^2\psi_I^\ast(r) \left( \frac{p^2}{2\mu}+ V(r)\right)  \psi_I(r)  dr+ \int_{r^{long}}^{r_0} r^2\psi_{II}^\ast(r) \left( \frac{p^2}{2\mu}+ V(r)\right)  \psi_{II}(r)  dr\right],
\end{equation}

where the wave functions $\psi_I(r)$ and $\psi_{II}(r)$ are as defined in equations (6) and (13) respectively.

\subsection{Slope and curvature of Isgur-Wise function}
Isgur, Wise, Georgi and others showed that in weak semi-leptonic decays of heavy-light mesons (e.g. $B$ mesons to $D$ or $D^\ast$ mesons), in the limit $m_Q\rightarrow \infty$ all the form factors that describe these decays are expressible in terms of a single universal function of velocity transfer, which is normalized to unity at zero-recoil. This function is known as the Isgur-Wise function. It measures the overlap of the wave functions of the light degrees of freedom in the initial and final mesons moving with velocities $v$ and $v\prime$ respectively.\\
 
The Isgur-Wise functions are denotd by $\xi(Y)$, where $Y=v.v^\prime$ and $\xi(Y)\vert_{Y=1}=1$ is the normalization condition at the zero-recoil point ( $v=v^\prime$ ) \cite{11}.\\

The calculation of Isgur-Wise function is non-perturbative in principle and is performed for different phenomenological wave functions for mesons \cite{5,6}. This function depends upon the meson wave function and some kinematic factor, as given below :\\
\begin{equation}
\xi(Y)=\int_0^{\infty} 4\pi r^2 \vert \psi(r)\vert^2 cos(pr)dr,
\end{equation}
where $\psi(r)$ is the wave function for light quark only and
\begin{equation}
cos(pr)=1- \frac{p^2 r^2}{2}+\frac{p^4 r^4}{24}+.....
\end{equation}
with $p^2=2\mu^2(Y-1)$.\\

Taking cos(pr) upto $O(r^4)$ we get,

\begin{equation}
\begin{split}
\xi(Y)= \int_0^{\infty} 4\pi r^2 \vert \psi(r)\vert^2dr-\left[ 4\pi \mu^2 \int_0^{\infty} r^4 \vert \psi(r)\vert^2 dr\right] (Y-1)+\\ \left[ \frac{2}{3}\pi \mu^4 \int_0^{\infty} r^6 \vert \psi(r)\vert^2 dr\right] (Y-1)^2.
\end{split}
\end{equation}

In an explicit form, the Isgur-Wise function can be written as \cite{iwf1,iwf}

\begin{equation}
\xi(Y)=1-\rho^2 (Y-1)+C(Y-1)^2 ,
\end{equation}

where $\rho^2>0$.\\

The quantity $\rho^2$ is the slope of the Isgur-Wise function which determines the behavior of Isgur-Wise function close to zero recoil point ($Y=1$) and known as charge radius:

\begin{equation}
\rho^2=\frac{\partial \xi}{\partial Y}\vert_{Y=1}.
\end{equation}

The second order derivative is the curvature of the Isgur-Wise function known as convexity parameter:
\begin{equation}
C=\frac{1}{2} \left( \frac{\partial^2 \xi}{\partial Y^2}\right) \vert_{Y=1}.
\end{equation}

A precise knowledge of the slope and curvature of $\xi(Y)$ basically determines the Isgur-Wise function in the physical region. In Heavy Quark Effective Theory (HQET) as proposed by Neubert \cite{iwf1}, the Isgur-Wise function at zero recoil point allows us to determine CKM element $|V_{cb}|$ \cite{ijmpa} for the semi leptonic decays $B^0\rightarrow D^*l\nu$ and $B^0\rightarrow Dl\nu$.\\

Now from equations (32) and (33),

\begin{equation}
\rho^2= 4\pi \mu^2 \int_0^{\infty} r^4 \vert \psi(r)\vert^2 dr,
\end{equation}
\begin{equation}
C=\frac{2}{3}\pi \mu^4 \int_0^{\infty} r^6 \vert \psi(r)\vert^2 dr
\end{equation}
and
\begin{equation}
\int_0^{\infty} 4\pi r^2 \vert \psi(r)\vert^2dr=1.
\end{equation}

In the present work, we improve the above equations for $\rho^2$ and $C$ to

\begin{equation}
\rho^2= 4\pi \mu^2 \left[ \int_0^{r^{short}} r^4 \vert \psi_I(r)\vert^2 dr + \int_{r^{long}}^{r_0} r^4 \vert \psi_{II}(r)\vert^2 dr\right] 
\end{equation}

and

\begin{equation}
C=\frac{2}{3} \pi \mu^4 \left[ \int_0^{r^{short}} r^6 \vert \psi_I(r)\vert^2 dr+\int_{r^{long}}^{r_0} r^6 \vert \psi_{II}(r)\vert^2 dr\right] .
\end{equation}

Using these modified expressions for slope and curvature of Isgur-Wise function in equation (33), we have computed the results. In equations (39) and (40), $\psi_I(r)$ and $\psi_{II}(r)$ are the wave functions as defined in (6) and (13) respectively.\\

Now to find the cut offs $r^{short}$ and $r^{long}$, we use the two choices of perturbative conditions:\\

choice-I: for Coulomb as parent and linear as perturbation
\begin{equation}
-\frac{4\alpha_s}{3r} \textgreater br
\end{equation}

and\\

choice-II: for linear as parent and Coulomb as perturbation

\begin{equation}
br \textgreater -\frac{4\alpha_s}{3r}.
\end{equation}

From (41) and (42) we can find the bounds on $r$ upto which choice-I and II are valid. Choice-I gives the cut off on the short distance $r_{max}^{short}<\sqrt{\frac{4\alpha_s}{3b}}$ and choice-II gives the cut off on the long distance $r_{min}^{long}>\sqrt{\frac{4\alpha_s}{3b}}$.\\

We make $r^{short}=r^{long}=\sqrt{\frac{4\alpha_s}{3b}}$ for our analysis, otherwise unless they are identical, the addition of two counterparts (linear part $\&$ Coulomb part) either overestimate or under estimate the calculated values of quantities which involves the integration over 0 to $r^{short}$ and $r^{long}$ to $r_0$ \cite{pramana}.\\

In table 1, we show the bounds on $r^{short}$ and $r^{long}$ in $Fermi$ which yields exact/most restrictive upper bounds of the quantities to be calculated. 

\begin{table}[h]
\caption{$r^{short}$ and $r^{long}$ in $Fermi$ with $c=0$ and $b=0.183 GeV^2$}
\begin{center}
\begin{tabular}{ll}
\hline
            $\alpha_s$-value           & $r^{short}=r^{long}$ \\
            &($Fermi$)\\ \hline
                 0.39 &  0.332\\     
            (for charmonium scale)   &\\ \hline
           
                    0.22 & 0.249\\
            (for bottomonium scale)         &    \\
  \hline
\end{tabular}
\end{center}
\end{table}

\section{Results}

We calculate the masses of various heavy-light mesons using equation (27) and the obtained results are compared with the experimental data \cite{k} in table 3. We have used Mathematica version 7.0.0 to compute the results.\\



The input parameters in the numerical calculations used are $m_u=0.336 GeV$, $m_s=0.483 GeV$, $m_c=1.55 GeV$, $m_b=4.95 GeV$ and $b=0.183 GeV^2$ and $\alpha_s$ values 0.39 and 0.22 for charmonium and bottomonium scale respectively, are same with the previous work \cite{ijp,pramana}.\\

With these values, the reduced masses ($\mu$) of the mesons, using equation (9) are shown in table 2. 

\begin{table}[!h]\footnotesize
\caption{Reduced masses of heavy-light mesons in $GeV$.}
\begin{center}
\begin{tabular}{ll}
\hline
   Meson              &  Reduced mass ($\mu$)\\
   & $(GeV)$  \\ \hline
                  
 $D(c\bar{u}/c\bar{d}$)   &   0.276  \\ 
   
 $D_s(c\bar{s})$          &  0.368\\
 
 $B(u\bar{b}/d\bar{b})$   & 0.314 \\ 
 
 $B_s(s\bar{b})$ &      0.440  \\    

 $B_c(\bar{b}c)$ &    1.180 \\ \hline

\end{tabular}
\end{center}
\end{table}

\begin{table}[h]\footnotesize
\caption{Masses of heavy-light mesons in GeV.}
\begin{center}
\begin{tabular}{llllll}
\hline
$\alpha_s$ & Meson & \multicolumn{1}{c}{$r^S=r^L$ ($Fermi$)} & Mass ($M_P$) $(GeV)$  & Experimental Mass $(GeV)$ \cite{k}\\  \hline

\multirow{2}{*}{0.39} & D(c$\bar{u}/c\bar{d}$) &  \multirow{2}{*}{0.332}  & 2.378 &         1.869$\pm$ 0.0016\\ 
   &      $D_s(c\bar{s})$&         & 2.500&1.968$\pm$ 0.0033\\  \cline{1-5} 
   
   \multirow{2}{*}{0.22} &  $B(u\bar{b}/d\bar{b})$ &\multirow{3}{*}{0.249} &5.798 & 5.279$\pm$ 0.0017\\
   & $B_s(s\bar{b})$ &&5.902& 5.366$\pm$ 0.0024\\
 & $B_c(\bar{b}c)$     & & 6.810 & 6.277$\pm$ 0.006\\ \hline

\end{tabular}
\end{center}
\end{table}

Our results for $B$ mesons are found to be more agreement with experimental data than $D$ mesons.\\ 

In table 4 and 5, we find slope ($\rho^2$ and $\rho^{\prime2}$) and curvature ($C$ and $C^\prime$) using modified equations (39) and (40) respectively.\\

The numerical results for $\rho^2$ and $C$ in the Isgur-Wise limit is shown in the table 4, where we consider the limit where the mass of active quark/anti-quark (in this case $b$-quark) is infinitely heavy ($m_Q/m_{\bar{Q}}\rightarrow \infty$) and the reduced mass $\mu$ becomes that of the light quark/anti-quark ($m_q/\bar{m_q}$) (in this case $u$-quark). We have also compared our results with the predictions of other theoretical models \cite{16,17,19,20,21}.\\

\begin{table}[h]\footnotesize
\caption{Values of $\rho^2$ and $C$ in the present work and other works in the limit $m_Q\rightarrow \infty$.}
\begin{center}
\begin{tabular}{llll}
\hline
                                            &   $\rho^2$     & $C$ \\ \hline
Present work                                &   1.176       & 0.180\\ \hline
Other work                                      &                & \\ \hline

Le Youanc et al. \cite{16}                             &$\geq$ 0.75     & 0.47\\
Rosner \cite{17}                                        & 1.66           & 2.76\\ 
Mannel \cite{19}                                     & 0.98           & 0.98\\
Pole Ansatz \cite{20}                                   & 1.42           & 2.71 \\
Ebert et al. \cite{21}                                  & 1.04           & 1.36\\ \hline

\end{tabular}
\end{center}
\end{table}

However, in a generalized way we can also check the flavor dependence of the form factor in heavy meson decays. We calculate the slope ($\rho^{\prime2}$) and curvature ($C^\prime$) of form factor of semi-leptonic decays in finite mass limit with the flavor dependent correction. In table 5, we compare our present results with the previous work \cite{5,14}. The results in the present work clearly shows an improvement of the previous analysis.

\begin{table}[h]\footnotesize
\caption{ Values of slope ($\rho^{\prime2})$ and curvature ($C^\prime$) of the form factor of heavy meson decays in the present work and previous work with finite mass correction }
\begin{center}
\begin{tabular}{llll}
\hline
                              & Meson              &   $\rho^{\prime2}$  & $C^\prime$ \\ \hline
\multirow{4}{*}{Present work}  & D(c$\bar{u}/c\bar{d}$)  & 0.911 & 0.106  \\ 
                             & $D_s(c\bar{s})$         &    1.318   & 0.228 \\   
                               & $B(u\bar{b}/d\bar{b})$  &      1.110    & 0.260 \\ 
                               & $B_s(s\bar{b})$         &   1.722   & 0.721 \\                      
                             &$B_c(c\bar{b})$         &  4.646       & 6.074\\ \hline
                              
\multirow{4}{*}{Previous work \cite{5,14}}  & D(c$\bar{u}/c\bar{d}$)  &1.136 & 5.377 \\ 
                             & $D_s(c\bar{s})$         &   1.083        &  3.583\\   
                               & $B(u\bar{b}/d\bar{b})$  &   128.28        &  5212\\ 
                               & $B_s(s\bar{b})$         &   112.759      &  4841\\ 
Previous work \cite{ijmpa}      &$B_c(c\bar{b})$         &  5.45         & 31.39 \\ \hline

\end{tabular}
\end{center}
\end{table}

\newpage
The variation of Isgur-Wise function $\xi(Y)$ with $Y$ in the Isgur-Wise limit is shown in figure-1(a) (using table 4), where the mass of the $b$-quark is considered to be infinitely heavy and the reduced mass $\mu$ is $0.336 GeV$ (mass of $u$ or $d$-quark/anti-quark). In a similar way, we draw the graph of figure-1(b) (using table 5) for finite mass and flavor dependent correction. Also for comparison the results of ref. \cite{17} and \cite{21} are plotted in both the graphs.\\

\begin{figure}[!h]
\subfloat[In the limit $m_Q\rightarrow \infty$]{%
  \includegraphics[width=0.45\linewidth]{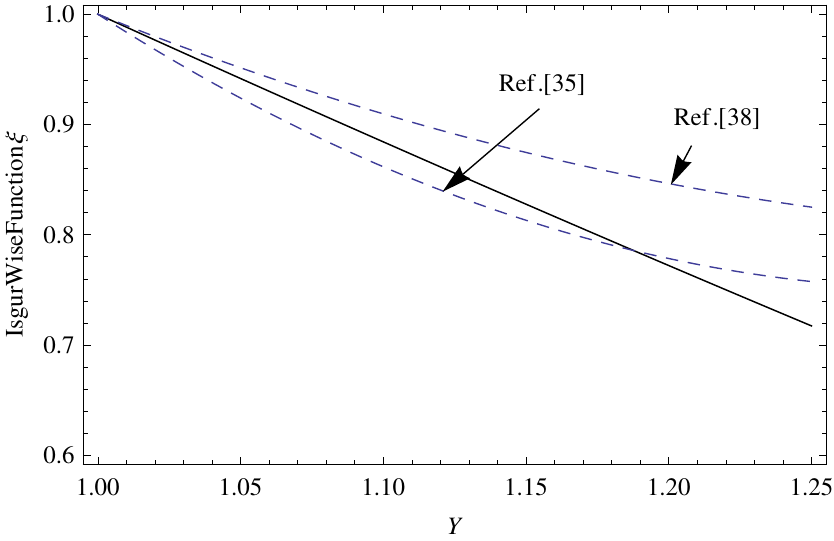}%
}\vspace{1ex}
\subfloat[Finite mass correction]{%
  \includegraphics[width=0.45\linewidth]{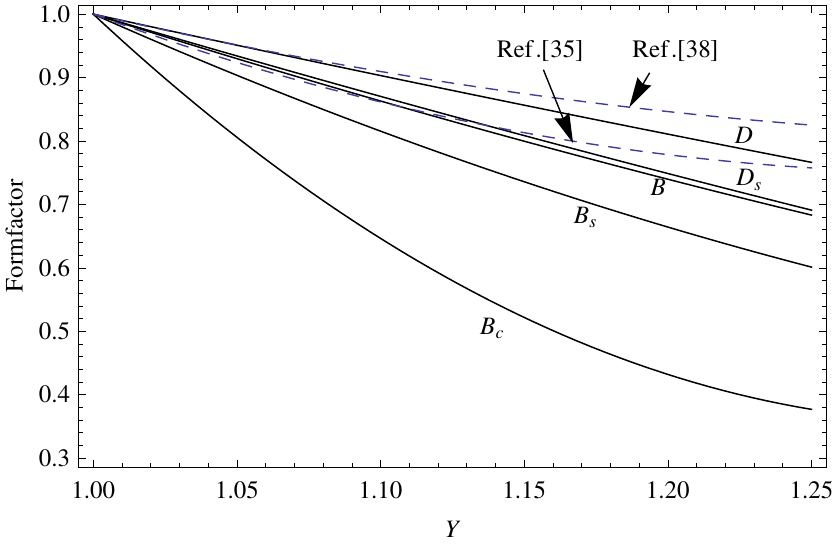}%
}\vspace{1ex}
\caption{Variation of form factor with $Y$ in the Isgur-Wise limit is represented in (a) and that of finite mass correction is represented in (b).}
\label{fig:1}
\end{figure}

To draw the graphs as shown in figure 1, we have used equations (39) and (40) in (33). $\xi(Y)$ is found to have expected fall with $Y=v.v^\prime$. It is also seen from the figure that the computed results are well within the other model values \cite{17,21}.

\section{Discussion and conclusion}
We have calculated the values of masses and convexity parameter of the Isgur-Wise function considering the scaling factor `$c$' as zero. One of the important point about this work is that we have given equal fitting to both the Coulomb and linear part of the Cornell potential unlike in the previous analysis \cite{kkbj,5,pd,13,bjdk,6,ijmpa,14}. Also our calculations provide a measure of the slope and curvature of the form factors with finite mass corrections. We can say that the modification induced by mass effect are not so significant. Furthermore, the consideration of the finite mass correction changes the results only slightly (significantly for $B(u\bar{b}/d\bar{b})$ meson). However, for the mesons where light quark/antiquark is not so light compared to the heavy quark/antiquark, the finite mass limit do show a very strong dependence on the spectator quark mass; for example we can see $B_c(c\bar{b})$ meson (table 5).\\

Our calculated values of masses of mesons are found to be in good agreement with the experimental datas (table 3). Also the calculated values of slope and curvature of Isgur-Wise function in this work are well within the limit of other theoretical values (table 4). However the re-evaluation of the model with a non-zero scaling factor with the satisfaction of the quantum mechanical idea is currently under study.\\

Let us conclude this paper with a comment that the relativity is by no means negligible for heavy-light systems. Such effects do not merely lead to a Dirac modification factor as used in the present work, but also have other significant effects as have been studied in various relativistic treatments of the problem \cite{21}. Inspite of the phenomenological success of the present model, it falls short of such expectation.

\section*{Acknowledgement}
\textit{The final version of the work was completed when one of us (DKC) was a visitor at the Rudolf Peierls Center for Theoretical Physics, University of Oxford, UK. He thanks Prof Subir Sarkar for his hospitality. One of the authors (TD) acknowledges the support of University Grants Commission in terms of fellowship under BSR scheme to pursue research work at Gauhati University, Department of Physics.}

\appendix
\numberwithin{equation}{section}
\appendixpageoff
\section{Appendix}

\section*{Wave function for Coulomb ($-\frac{4\alpha_s}{3r})$ as parent and linear (br) as perturbation:}

The first order perturbed eigenfunction $\psi^{(1)}$ and first order eigen energy $W^{(1)}$ using quantum mechanical perturbation theory (Dalgarno's method) can be obtained using the relation
\begin{equation}	
H_0 \psi^1 + H^\prime \psi^{(0)}=W_0\psi^{(1)} + W^{(1)} \psi^{(0)},
\end{equation}

where
\begin{equation}
W^{(1)}= <\psi^{(0)}\vert H^\prime \vert \psi^{(0)}>
\end{equation}

\begin{equation}
=\int \psi^\star_{100} H^\prime \psi_{100} d\tau.
\end{equation}

For Cornell potential (5), we consider

\begin{equation}
H_0=-\frac{4\alpha_s}{3r}
\end{equation}

and

\begin{equation}
H^\prime=br.
\end{equation}

From (A.1) we obtain
\begin{equation}
(H_0-W^{(0)})\psi^{(1)}=(W^{(1)}-H^\prime)\psi^{(0)}.
\end{equation}

Putting
\begin{equation}
A=\frac{4 \alpha_s}{3},
\end{equation}

we obtain
\begin{equation}
H_0=-\frac{\hbar^2}{2\mu}  \nabla^2-\frac{A}{r},
\end{equation}

\begin{equation}
W^{(0)}=E=\frac{\mu A^2}{2}
\end{equation}

and
\begin{equation}
\psi^{(0)}(r)=\frac{1}{\sqrt{\pi a_{0}^3}}e^{-\frac{r}{a_0}},
\end{equation}

where $\psi^{(0)}$ is the unperturbed wave function in the zero$^{th}$ order of perturbation and $a_0$ is given by equation (8).\\

Taking $\hbar^2=1$, equation (A.6),
\begin{equation}
\Rightarrow \left( -\frac{\hbar^2}{2\mu}\nabla^2-\frac{A}{r}-E\right) \psi^{(1)}
=\left( W^{(1)}-br\right) \frac{1}{\sqrt{\pi a_{0}^3}}e^{-\frac{r}{a_0}} \nonumber
\end{equation}

\begin{equation}
\Rightarrow \left( \nabla^2+\frac{2\mu A}{r} -\mu^2 A^2\right) \psi^{(1)}=(br-W^{(1)})\frac{2\mu}{\sqrt{\pi a_{0}^3}}e^{-\frac{r}{a_0}} \nonumber
\end{equation}

\begin{equation}
\Rightarrow \left( \nabla^2+\frac{2}{a_0 r}-\frac{1}{a_0^2}\right) \psi^{(1)}=(br-W^{(1)})\frac{2\mu}{\sqrt{\pi a_{0}^3}}e^{-\frac{r}{a_0}}.
\end{equation}

Let
\begin{equation}
\psi^{(1)}=(br)R(r),
\end{equation}

then

\begin{equation}
(A.11)\Rightarrow \left( \frac{d^2}{dr^2}+\frac{2}{r}\frac{d}{dr}+\frac{2}{a_0r}-\frac{1}{a_0^2}\right) \left( br\right) R(r)=D(br-W^{(1)})e^{-\frac{r}{a_0}},
\end{equation}

where we put
\begin{equation}
D=\frac{2\mu}{\sqrt{\pi a_{0}^3}}.
\end{equation}

Now
\begin{equation}
\frac{d}{dr}\left(brR(r)\right) =bR(r)+br\frac{dR}{dr},
\end{equation}

\begin{equation}
\frac{d^2}{dr^2}\left( brR(r)\right) =2b\frac{dR}{dr}+br\frac{d^2R}{dr^2}
\end{equation}

Using (A.15) and (A.16) in (A.13) we obtain

\begin{equation}
\begin{split}
br\frac{d^2R}{dr^2}+2b\frac{dR}{dr}+\frac{2}{r}bR(r)+\frac{2}{r}br\frac{dR}{dr}+\frac{2}{a_0r}brR(r)-\frac{1}{a_0^2}brR(r)\\=D(br-W^{(1)})e^{-\frac{r}{a_0}}.
\end{split}
\end{equation}

Putting
\begin{equation}
R(r)=F(r)e^{-\frac{r}{a_0}},
\end{equation}

\begin{equation}
\frac{dR}{dr}=F^\prime(r) e^{-\frac{r}{a_0}} - \frac{1}{a_0}F(r)e^{-\frac{r}{a_0}},
\end{equation}

\begin{equation}
\frac{d^2R}{dr^2}=F^{\prime\prime}(r)e^{-\frac{r}{a_0}}-\frac{2}{a_0}F^\prime (r)e^{-\frac{r}{a_0}}+\frac{1}{a_0^2}F(r)e^{-\frac{r}{a_0}},
\end{equation}

\begin{equation}
\begin{split}
(A.17)\Rightarrow br\left\lbrace F^{\prime\prime}(r)-\frac{2}{a_0}F^\prime (r)+\frac{1}{a_0^2}F(r)\right\rbrace +2b\left\lbrace F^\prime(r)-\frac{1}{a_0}F(r)\right\rbrace + \frac{2b}{r}F(r)\\+2b F^\prime(r)-\frac{2b}{a_0}F(r)+\frac{2b}{a_0}F(r)-\frac{1}{a_0^2}brF(r)=D(br-W^{(1)}) \nonumber
\end{split}
\end{equation}

\begin{equation}
\Rightarrow brF^{\prime\prime}(r)+\left\lbrace 4b-\frac{2b}{a_0}r\right\rbrace F^\prime(r)+\left\lbrace \frac{2b}{r}-\frac{2b}{a_0}\right\rbrace F(r)=D(br-W^{(1)}).
\end{equation}

Let
\begin{equation}
F(r)=\sum_{n=0}^\infty A_nr^n,
\end{equation}

then
\begin{equation}
F^\prime(r)=\sum_{n=0}^\infty nA_nr^{n-1}
\end{equation}

and
\begin{equation}
F^{\prime\prime}(r)=\sum_{n=0}^\infty n(n-1)A_nr^{n-2}.
\end{equation}

\begin{equation}
\begin{split}
(A.21)\Rightarrow br\sum_{n=0}^\infty n(n-1)A_nr^{n-2}+\left\lbrace 4b-\frac{2b}{a_0}r\right\rbrace \sum_{n=0}^\infty nA_nr^{n-1}+ \left\lbrace \frac{2b}{r}-\frac{2b}{a_0}\right\rbrace \sum_{n=0}^\infty A_nr^n\\=D(br-W^{(1)}) \nonumber
\end{split}
\end{equation}

\begin{equation}
\begin{split}
\Rightarrow \left\lbrace b\sum_{n=0}^\infty n(n-1)A_n+4b\sum_{n=0}^\infty nA_n+2b\sum_{n=0}^\infty A_n\right\rbrace r^{n-1}- \left\lbrace \frac{2b}{a_0}\sum_{n=0}^\infty nA_n+\frac{2b}{a_0}\sum_{n=0}^\infty A_n\right\rbrace r^n\\=D(br-W^{(1)}).
\end{split}
\end{equation}

Equating the coefficients of $r^{-1}$ on both sides of the above identity (A.25)
\begin{equation}
2bA_0=0 \nonumber,
\end{equation}

since $b\neq0$, therefore 

\begin{equation}
\Rightarrow A_0=0.
\end{equation}

Equating the coefficients of $r^0$ on both sides of the identity (A.25),
\begin{equation}
4bA_1+2bA_1-\frac{2b}{a_0}A_0=-DW^{(1)} \nonumber
\end{equation}

\begin{equation}
\Rightarrow A_1=-\frac{DW^{(1)}}{6b}.
\end{equation}

Equating the coefficients of $r^1$ on both sides of the identity (A.25),
\begin{equation}
2bA_2+8bA_2+2bA_2-\frac{2b}{a_0}A_1-\frac{2b}{a_0}A_1=Db \nonumber.
\end{equation}

Using(A.27) and (A.26),
\begin{equation}
A_2=\frac{D}{12}-\frac{D W^{(1)}}{18ba_0} .
\end{equation}

Equating the coefficients of $r^2$ on both sides of the identity (A.25),
\begin{equation}
6bA_3+12bA_3+2bA_3-\frac{4b}{a_0}A_2-\frac{2b}{a_0}A_2=0.
\end{equation}

Using (A.27) and (A.28),
\begin{equation}
A_3=\frac{D}{40a_0}-\frac{DW^{(1)}}{60ba_0^2} .
\end{equation}

Equating the coefficients of $r^3$ on both sides of the identity (A.25),
\begin{equation}
12bA_4+16bA_4+2bA_4-\frac{2b}{a_0}3A_3-\frac{2b}{a_0}A_3=0. \nonumber
\end{equation}

Using (A.28) and (A.30),
\begin{equation}
A_4=\frac{D}{150a_0^2}-\frac{DW^{(1)}}{225ba_0^3}.
\end{equation}

From (A.22)
\begin{equation}
F(r)=A_0r^0+A_1r^1+A_2r^2+A_3r^3+A_4r^4+...
\end{equation}

Now from (A.12), (A.18) and (A.32),

\begin{equation}
\psi^{(1)}(r)=brF(r)e^{-\frac{r}{a_0}}
\end{equation}

\begin{equation}
=br\left( A_0r^0+A_1r^1+A_2r^2+A_3r^3+A_4r^4+...\right)e^{-\frac{r}{a_0}} \nonumber
\end{equation}

\begin{equation}
=\left\lbrace A_0(br)+A_1(br^2)+A_2(br^3)+A_3(br^4)+A_4(br^5)+...\right\rbrace e^{-\frac{r}{a_0}} .
\end{equation}

Now applying (A.26),(A.27),(A.28),(A.30),(A.31) to (A.34),

\begin{equation}
\begin{split}
\psi^{(1)}(r)=\biggl[-\frac{DW}{6b}(br^2)+\left\lbrace \frac{D}{6}\left( \frac{1}{2}-\frac{W}{3ba_0}\right)\right\rbrace (br^3)
 +\left\lbrace\frac{D}{20a_0}\left( \frac{1}{2}-\frac{W}{3ba_0}\right)\right\rbrace (br^4)\\ 
 +\left\lbrace\frac{D}{75a_0^2}\left( \frac{1}{2}-\frac{W}{3ba_0}\right)\right\rbrace (br^5)\biggr]e^{-\frac{r}{a_0}}  .
\end{split}
\end{equation}

Again from (A.2)

\begin{equation}
W^{(1)}=\int \psi^\star_{100} H^\prime \psi_{100} d\tau \nonumber
\end{equation}

\begin{equation}
=\frac{1}{\pi a_0^3}\int_0^{\infty}(br)r^2 e^{-\frac{2r}{a_0}} dr \int_0^{\pi}Sin\theta d\theta \int_0^{2\pi}d\phi \nonumber
\end{equation}

\begin{equation}
=\frac{4\pi}{\pi a_0^3}\int_0^{\infty}(br^3) e^{-\frac{2r}{a_0}} dr \nonumber
\end{equation}

\begin{equation}
=\frac{4}{a_0^3}\left[ b\frac{6a_0^4}{16}\right] \nonumber
\end{equation}

\begin{equation}
=\frac{3}{2}ba_0.
\end{equation}

Hence
\begin{equation}
\frac{1}{2}-\frac{W}{3ba_0}=0.
\end{equation}

Therefore, (A.35) reduces to
\begin{equation}
\begin{split}
\psi^{(1)}(r)=\left[-\frac{DW}{6b}(br^2)\right]e^{-\frac{r}{a_0}} \nonumber
\end{split}
\end{equation}

\begin{equation}
=-\frac{1}{2\sqrt{\pi a_0^3}}\mu b a_0 r^2 e^{-\frac{r}{a_0}}.
\end{equation}

The total wave function is thus
\begin{equation}
\psi^{total}=\psi^{(0)}+\psi^{(1)} \nonumber
\end{equation}

\begin{equation}
=\frac{1}{\sqrt{\pi a_0^3}}\left[ 1-\frac{1}{2}\mu b a_0r^2\right] e^{-\frac{r}{a_0}}.
\end{equation}

Considering relativistic effect the above equation becomes
\begin{equation}
\psi^{total}(r)=\frac{N}{\sqrt{\pi a_0^3}}\left[1-\frac{1}{2}\mu b a_0r^2\right]\left( \frac{r}{a_0}\right) ^{-\epsilon} e^{-\frac{r}{a_0}}.
\end{equation}

\section{Appendix}
\section*{Wave function for linear ($br$) as parent and Coulomb ($-\frac{4\alpha_s}{3r}$) as perturbation}
\numberwithin{equation}{section}

Here we take $br$ as parent and $-\frac{4\alpha_s}{3r}$ as perturbation so that 

\begin{equation}
H_0=-\frac{\hbar^2}{2\mu} \nabla^2+br
\end{equation}

with
\begin{equation}
H^\prime=-\frac{4\alpha_s}{3r}.
\end{equation}


To find the unperturbed wave function corresponding to $H_0$ we employ the radial Schrodinger equation for potential $br$ for ground state,

\begin{equation}
-\frac{1}{2\mu}\left[ \left( \frac{d^2}{dr^2}+\frac{2}{r}\frac{d}{dr}\right)+br\right]  R(r)=ER(r),
\end{equation}

where $R(r)$ is the radial wave function. We introduce $u(r)=rR(r)$ and the dimensionless variable

\begin{equation}
\rho(r)=(2\mu b)^{\frac{1}{3}}r-\left( \frac{2\mu}{b^2}\right) ^{\frac{1}{3}}E.
\end{equation}

The equation (B.3) then reduces to
\begin{equation}
\frac{d^2u}{d\rho^2}-\rho u=0.
\end{equation}

The solution of this second order homogeneous differential equation contains linear combination of two types of Airy's functions $Ai[r]$ and $Bi[r]$. The nature of the Airy's function reveals that

\begin{center}
$Ai[r]\rightarrow 0$ and $Bi[r]\rightarrow \infty$ as $r\rightarrow \infty$.
\end{center}

So, it is reasonable to reject the $Bi[r]$ part of the solution.\\

The unperturbed wave function \cite{6} for ground state is

\begin{equation}
\psi^{(0)}(r)=\frac{N_0}{r}Ai[\rho_1r+\rho_0],
\end{equation}

where $N_0$ is the normalization constant and $\rho_1=(2\mu b)^{1/3}$ .\\

$\rho_0$ is the zero of the Airy function, such that $Ai[\rho_0]=0$.\\

$\rho_0$ has the explicit form as mentioned in equation (21).\\

The first order perturbed eigen function $\psi^{(1)}$ can be calculated using relation (A.6).\\

Then taking $\hbar^2=1$, equation (A.6),
\begin{equation}
\Rightarrow \left( -\frac{\hbar^2}{2\mu}\nabla^2+br-E\right) \psi^{(1)}
=\left( W^{(1)}+\frac{4\alpha_s}{3r}\right) \psi^{(0)}(r) .
\end{equation}

In terms of the radial wave function the above equation can be expressed as
\begin{equation}
\left[ \left( \frac{d^2}{dr^2}+\frac{2}{r}\frac{d}{dr}\right) -2\mu(br-E)\right]  R(r)
=-2\mu\left( W^{(1)}+\frac{4\alpha_s}{3r}\right)\frac{1}{r}Ai[\rho].
\end{equation}

Let 
\begin{equation}
R(r)=\frac{1}{r}F(r)Ai[\rho]=\frac{1}{r}F(r)Ai[\rho_1r+\rho_0],
\end{equation}

so that
\begin{equation}
\frac{dR}{dr}=-\frac{1}{r^2}F(r)Ai[\rho]+\frac{1}{r}F^\prime(r)Ai[\rho]+\frac{\rho_1}{r}F(r)Ai^\prime[\rho],
\end{equation}

\begin{equation}
\begin{split}
\frac{d^2R}{dr^2}=\frac{2}{r^3}F(r)Ai[\rho]-\frac{2}{r^2}F^\prime(r)Ai[\rho]-\frac{2\rho}{r^2}F(r)Ai^\prime[\rho]+\frac{1}{r}F^{\prime\prime}Ai[\rho_1]+\\ \frac{2\rho_1}{r}F^\prime(r)Ai^\prime[\rho]+\frac{\rho_1^2}{r}F(r)Ai^{\prime\prime}[\rho].
\end{split}
\end{equation}

Now we introduce the identity
\begin{equation}
Ai^\prime[\rho]=\frac{dAi(\rho)}{dr}=Z(\rho)Ai(\rho),
\end{equation}

so that
\begin{equation}
Ai^{\prime\prime}(\rho)=Z^2(\rho)Ai(\rho)+Z^\prime(\rho)Ai(\rho).
\end{equation}

Then the equation (B.8) becomes

\begin{equation}
\begin{split}
F^{\prime\prime}(r)+2\rho_1F^\prime(r)Z(\rho)+\rho_1^2[Z^2(\rho)+Z^\prime(\rho)]F(r)-2\mu(br-E)F(r)\\=-\frac{4\alpha_s}{3}\frac{2\mu}{r}-2\mu W^{(1)}.
\end{split}
\end{equation}

Assuming
\begin{equation}
Z(\rho)=\frac{k_1(r)}{r} \nonumber
\end{equation}

and
\begin{equation}
Z^2(\rho)+Z^\prime(\rho)=\frac{k_2(r)}{r^2}, \nonumber
\end{equation}

\begin{equation}
(B.14)\Rightarrow F^{\prime\prime}(r)+2\rho_1F^{\prime}(r)\frac{k_1(r)}{r}+\rho_1^2F(r)\frac{k_2(r)}{r^2}-2\mu(br-E)F(r)=-\frac{4\alpha_s}{3}\frac{2\mu}{r}-2\mu W^{(1)}.
\end{equation}

Now using (A.22), (A.23) and (A.24), the above equation (B.15) becomes

\begin{equation}
\begin{split}
n(n-1)\sum_nA_nr^{n-2}+2\rho_1l\sum_nA_nr^{n-1}\frac{k_1}{r}+\rho_1^2\sum_nA_nr^n\frac{k_2}{r^2}-2\mu(br-E)\sum_nA_nr^n\\=-\frac{4\alpha_s}{3}\frac{2\mu}{r}-2\mu W^{(1)}
\end{split}
\end{equation}

\begin{equation}
\begin{split}
\Rightarrow \left[n(n-1)\sum_nA_n+ 2\rho_1n\sum_nA_nk_1+\rho_1^2\sum_nA_nk_2\right]r^{n-2}-2\mu b\sum_nA_nr^{n+1}+
\\2\mu E\sum_nA_nr^n=-\frac{4\alpha_s}{3}\frac{2\mu}{r}-2\mu W^{(1)}.
\end{split}
\end{equation}

Now equating the co-efficients of $r^{-2}$ from the above equation (B.17),

\begin{equation}
\rho_1^2A_0k_2=0 \nonumber
\end{equation}

\begin{equation}
\Rightarrow A_0=0.
\end{equation}

Equating the co-efficients of $r^{-1}$ of (B.17),
\begin{equation}
2\rho_1A_1k_1+\rho_1^2A_1k_2=-2\mu\frac{4\alpha_s}{3}, \nonumber
\end{equation}

\begin{equation}
\Rightarrow A_1=\frac{-2\mu \frac{4\alpha_s}{3}}{2\rho_1 k_1+\rho_1^2 k_2}.
\end{equation}

Equating the co-efficients of $r^{0}$ of (B.17),

\begin{equation}
2A_2+4\rho_1A_2k_1+\rho_1^2A_2k_2+2\mu EA_0=-2\mu W^{(1)} \nonumber
\end{equation}

\begin{equation}
\Rightarrow A_2=\frac{-2\mu W^{(1)}}{2+4 \rho_1 k_1+ \rho_1^2 k_2}.
\end{equation}

Equating the co-efficients of $r^{1}$ of (B.17),

\begin{equation}
6A_3+6\rho_1A_3k_1+\rho_1^2A_3k_2-2\mu b A_0+2\mu EA_1=0 \nonumber
\end{equation}

\begin{equation}
\Rightarrow A_3=\frac{-2\mu EA_1}{6+6 \rho_1 k_1+ \rho_1^2 k_2}.
\end{equation}

Equating the co-efficients of $r^2$ of (B.17),
\begin{equation}
12A_4+8\rho_1A_4k_1+\rho_1^2A_4k_2-2\mu b A_1+ 2\mu EA_2=0 \nonumber
\end{equation}

\begin{equation}
\Rightarrow A_4=\frac{-2\mu EA_2+2\mu b A_1}{12+8 \rho_1 k_1+ \rho_1^2 k_2}.
\end{equation}

Using (A.32), the perturbed wave function will be
\begin{equation}
\psi^{(1)}(r)=\frac{1}{r}[A_0r^0+A_1r^1+A_2r^2+A_3r^3+A_4r^4+...]Ai[\rho_1r+\rho_0].
\end{equation}

Now considering upto $O(r^4)$ with relativistic effect the total wave function is thus
\begin{equation}
\psi^{total}(r)=\frac{N^\prime}{r}[1+A_0r^0+A_1r^1+A_2r^2+A_3r^3+A_4r^4]Ai[\rho_1r+\rho_0]\left( \frac{r}{a_0}\right) ^{-\epsilon}.
\end{equation}

\end{document}